\documentclass{PoS_arXiv}
\usepackage{amsmath}

\newcommand{\SM}{{\text{SM}}}
\newcommand{\MOD}{{\text{model}}}

\newcommand{\MIN}{{\text{min}}}
\newcommand{\MAX}{{\text{max}}}
\newcommand{\Br}{{\text{BR}}}

\newcommand{\HB}{{\tt HiggsBounds}}
\newcommand{\vers}{{\tt 2}.{\tt 0}.{\tt 0}}
\newcommand{\firstvers}{{\tt 1}.{\tt 0}.{\tt 0}}
\newcommand{\OBS}{{\text{obs}}}
\newcommand{\EXPEC}{{\text{expec}}}

\newcommand{\gev}{\text{GeV}}
\newcommand{\tev}{\text{TeV}}

\title{Introducing HiggsBounds 2.0.0}

\ShortTitle{Introducing HiggsBounds 2.0.0}

\author{Philip Bechtle\\
        DESY, Notkestrasse 85, 22607 Hamburg, Germany
	}
\author{\speaker{Oliver Brein}\\
        Physikalisches Institut,
	Albert-Ludwigs-Universit\"at Freiburg,\\
	Hermann-Herder-Str. 3, D-79104 Freiburg im Breisgau, Germany
	}
\author{Sven Heinemeyer\\
        Instituto de Fisica de Cantabria (CSIC-UC),
	Santander, Spain
	}
\author{Georg Weiglein\\
        DESY, Notkestrasse 85, 22607 Hamburg, Germany
	}
\author{Karina Williams\\
        Bethe Center for Theoretical Physics,
	Physikalisches Institut der
	Universit\"at Bonn,\\
	Nussallee 12, 53115 Bonn, Germany
	}

\abstract{
We introduce version 2.0.0 of the computer program {\tt
HiggsBounds}\footnote{
For code download, online version, updated manual
see : {\tt www.ippp.dur.ac.uk/HiggsBounds/}}.
The program 
tests neutral and
charged Higgs sectors of arbitrary models against the current exclusion bounds
from %
LEP and the Tevatron.  As input, it requires a
selection of model predictions, such as Higgs masses, branching ratios,
effective couplings and total decay widths. 
The program
uses the expected and observed topological cross section limits from the
Higgs searches to determine 
whether a given parameter scenario of a model is 
excluded at the 95\% C.L. by those searches.
Version 2.0.0 includes 39/53 LEP/Tevatron Higgs search analyses, 
compared to 13/36 in the previous release (1.2.0).
Among the newly included analyses are 
LEP searches for neutral Higgs bosons ($H$) decaying invisibly or into hadrons,
LEP searches via the production modes $\tau^+\tau^- H$ and $b\bar b H$,
Tevatron searches via $t\bar t H$, 
and LEP and Tevatron searches for charged Higgs bosons.
Also, all 
Tevatron results presented at the 
ICHEP'10 are included in version 2.0.0.
In this note, we explain the basic ideas behind 
the implementation of {\tt HiggsBounds}
and provide a list of search topologies implemented in
version 2.0.0.
Furthermore, we apply {\tt HiggsBounds} 
\vers\ to
(a) determine the allowed Higgs mass range for a simple 4th generation model,
(b) update/reproduce LEP/Tevatron Higgs exclusion plots 
for the MSSM $m_h^{\text{max}}$ benchmark
scenario,
and
(c) show exclusion results for the scalar sector of the Randall-Sundrum
model.
}

\FullConference{Third International Workshop on Prospects for Charged Higgs Discovery at Colliders - CHARGED2010,\\
		September 27-30, 2010\\
		Uppsala Sweden}

\begin{document}

\section{Motivation}
The search for Higgs bosons, neutral or charged, is a major part 
in the endeavour to unravel the nature of electroweak symmetry
breaking. 
The discovery of a charged Higgs boson, in particular, would
be an unambiguous sign of an extended Higgs sector.
The LEP experiments have 
searched for the Standard Model (SM) Higgs boson~\cite{LEP-SM-Higgs-analysis}
and many 
others,
and so do the Tevatron and LHC experiments.
So far, no signals of Higgs bosons have been found, and 
LEP \& Tevatron experimentalists 
turned this non-observation 
into cross section constraints.
The constraints 
are provided 
in the form of limits on cross sections of 
individual signal topologies (such as $e^+e^-\to h_iZ\to b\bar{b}Z$
or $p\bar p\to h_i Z\to b\bar b l^+l^-$) or in the form of combined limits for a
specific model, such as the SM. In the latter case,
the individual topological cross sections have been combined assuming the
proportions of the individual 
contributions to be as predicted by the model. 
\begin{figure}[b]
\includegraphics[width=.33\textwidth]{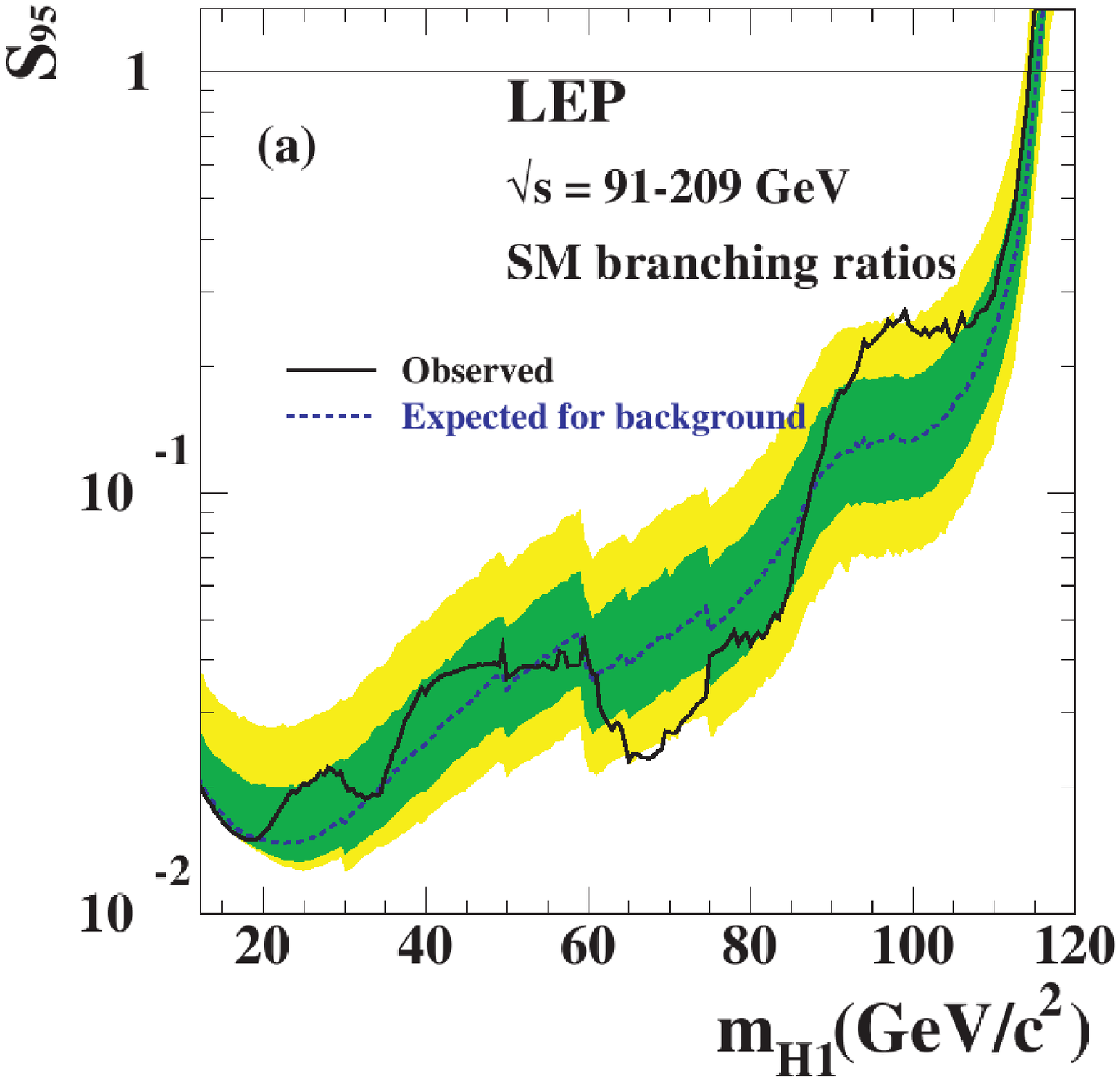}
\includegraphics[width=.33\textwidth]{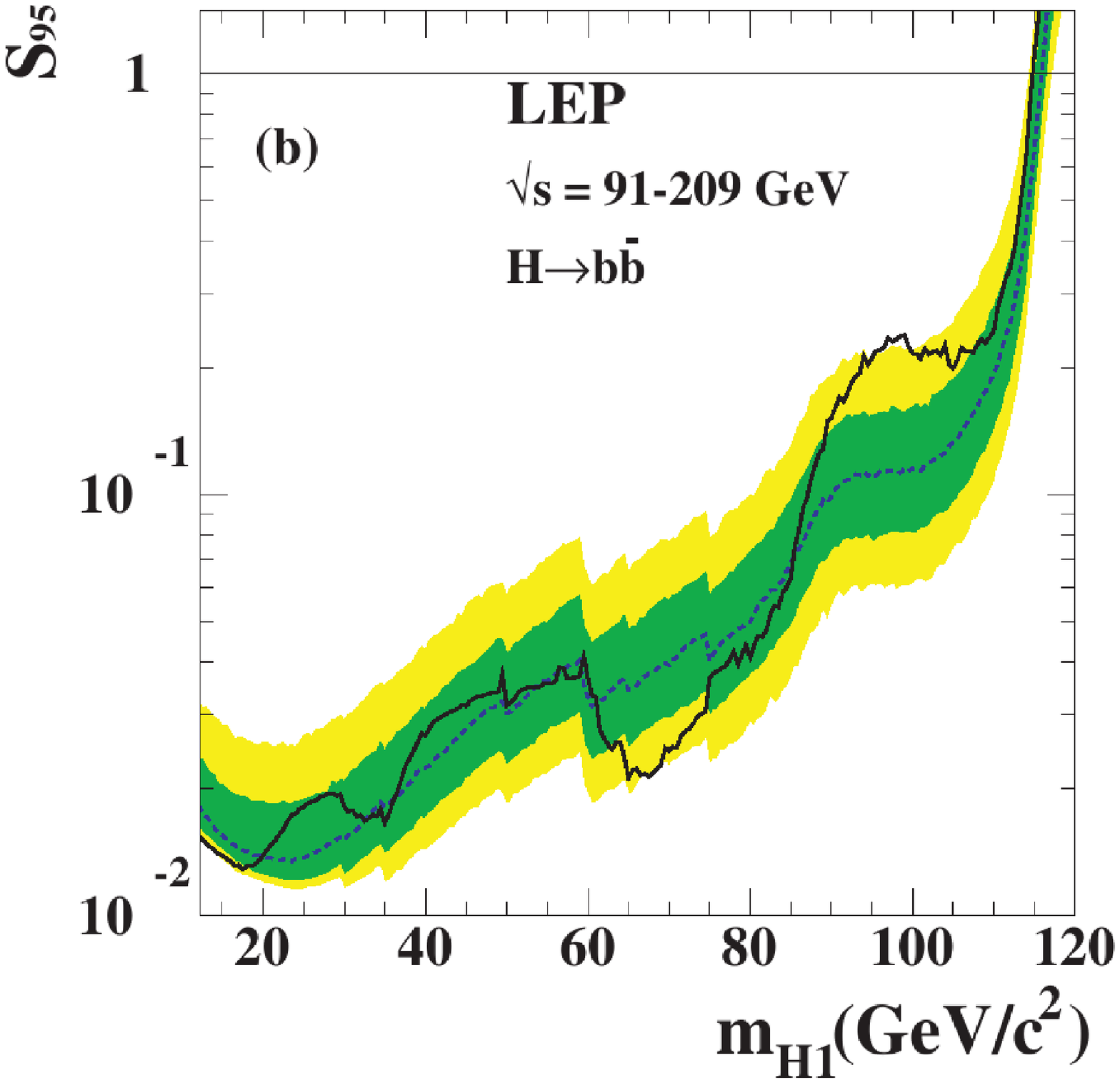}
\includegraphics[width=.33\textwidth]{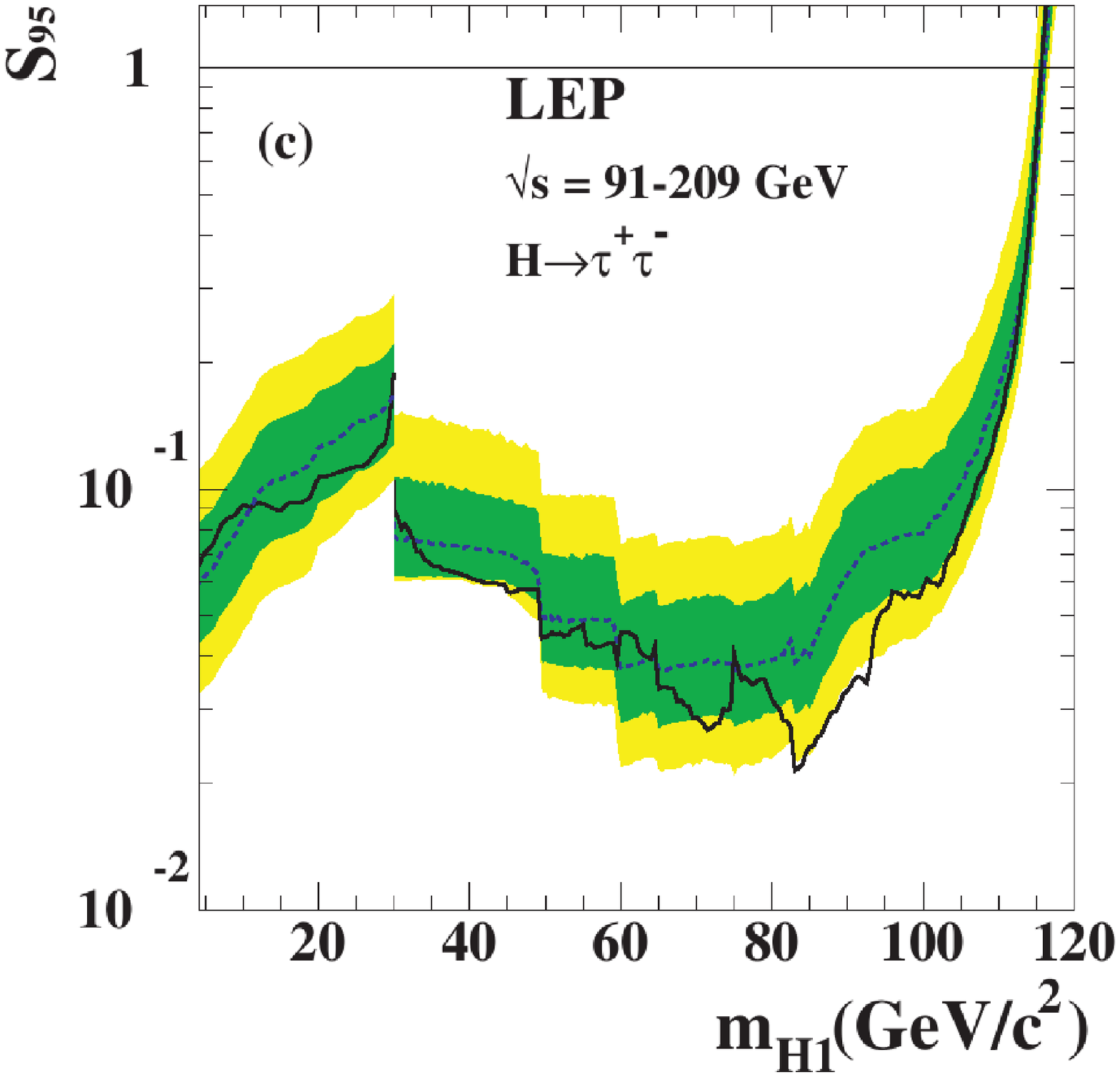}
\caption{\label{fig1}
Upper bound on the expected (dashed) and observed (solid) cross section
ratio $S_{95}$ for Higgs production via Higgsstrahlung, assuming
for the Higgs decay:
(a) SM branching ratios, (b) 100\% decay into $b\bar b$, 
or (c) 100\% decay into $\tau^+\tau^-$. \cite{LEP-MSSM-Higgs-analysis}
}
\end{figure}

The claim of model exclusion using non-observation constraints
is a statement about the statistical rejection of the 
``Higgs hypothesis'' (i.e. that Higgs signal + background describe the data).
Exclusion at the 95\% C.L. means that the Higgs hypothesis
has at most 5\% probability.
If no Higgs signal has been found, 
the lower the predicted signal cross section, the more probable 
is the Higgs hypothesis.
Hence, for given mass $m_{H1}$ and search channel for the Higgs boson,
one can determine
a lower bound $\sigma_\MIN(m_{H1})$, %
for cross sections above which the probability 
of the Higgs hypothesis stays below 5\%.
A model with cross section $\sigma_\MOD(m_{H1})>\sigma_\MIN(m_{H1})$
for the given search channel is then said to be excluded at the 
95\% C.L. 
Examples of typical (here: LEP) search results are shown in Fig.~\ref{fig1}.
Each of them assumes a particular Higgs decay pattern:
(a) SM decay branching ratios,
(b) 100\% decay into $b\bar b$,
and (c) 100\% decay into $\tau^+\tau^-$.
As the decay pattern is fixed, the limit is given on the 
Higgs production cross section (Higgsstrahlung) only (and normalised 
to the SM prediction): $S_{95}(m_{H1}) := 
        {\displaystyle \sigma^{HZ}_\MIN (m_{H1})}/
             {\displaystyle \sigma^{HZ}_\SM (m_{H1})}$.

For the SM result in Fig.~\ref{fig1}(a), several individual Higgs 
search topologies have been combined in order to increase the 
sensitivity to a SM Higgs boson.
Hence, the result applies only to SM-like models, i.e. models
where all individual signal cross sections differ from
the SM at most by the multiplication with one common
proportionality factor. 
Fig.~\ref{fig1}(b) and Fig.~\ref{fig1}(c) show results for 
single search topologies, which can be applied to any
model which predicts Higgs production via Higgsstrahlung
and has a non-zero BR to $b\bar b$ or $\tau^+\tau^-$, respectively.
If for a model which is not SM-like
$\sigma^{HZ}_\MOD (m_{H}) \Br_\MOD(H\to b\bar
b)/\sigma^{HZ}_\SM (m_{H1}) > S_{95}^{\text{observed}}(m_H)$ in Fig.~\ref{fig1}(b),
then the model is excluded at the 95\% C.L. by this search topology.

If exclusion results on several individual search topologies are available 
for a given model, the question arises {\em which} one to apply 
in order to determine 
if it is excluded.
Claiming exclusion of the model if at least one search topology
excludes it certainly does not preserve the 95\% C.L. 
and, therefore, should not be called exclusion.
The way out 
is to always consider only one
search result for excluding a model, but chose 
the analysis with the highest statistical sensitivity
for that. 
This information is 
encoded in
the expected limit
(dashed lines in Fig.~\ref{fig1}), the limit which would
result if only background events were present in the data
(obtained via Monte Carlo simulation).

\section{Implementation}

\HB\  tests the predictions of 
models with arbitrary Higgs sectors against
exclusion bounds
obtained 
from LEP/Tevatron Higgs searches.
It provides 
access to all relevant Higgs exclusion limits,
including expected limits.
As input, the program requires the predictions of the considered model for:
the number of neutral and singly %
charged Higgs bosons $h_i$ , $m_{h_i}$,
Higgs total widths $\Gamma_{\text{tot}}(h_i)$,
decay branching ratios BR$(h_i \to \ldots)$,
and production cross section ratios with respect to reference values.
For a given model scenario, 
\HB\ considers all implemented LEP and Tevatron analyses, 
checks whether they can be applied to the model considered,
and returns whether the scenario is excluded
at the 95\%\ C.L. or not. 
There are three ways to use \HB: a 
command line version, a subroutine version (written in Fortran 77 and 90)
a web interface.

The Tevatron and LEP cross section limits are understood 
to be applicable to models which do not 
change the signature of the background processes considerably
and are usually 
given for a narrow-width Higgs boson.
From this and other sources result some limitations on 
the applicability of \HB\ which are described 
in detail in \cite{HB100}.

Now to a 
crucial bit of the code,
namely, how one can make a statement at the 95\%\ C.L.
while considering several search analyses in parallel.
First a definition:
we call an ``analysis application'' $X$,
the application of a Higgs search analysis to a
particular Higgs boson (or e.g. {\em two} Higgs bosons, 
if the analysis involves {\em two} of them) of the model under 
study with particular mass(es).
To each analysis application $X$ corresponds a signal cross section 
prediction $\sigma(X)$ for the particular Higgs boson
on which an upper limit is put by the analysis.
For each $X$, {\tt HiggsBounds} uses the input to calculate
the quantity $Q_\MOD(X)$, which is %
$\sigma(X)$ up to a normalisation factor. 
In order to ensure the correct statistical interpretation of the results, it
is crucial to only consider the experimentally observed limit for one
particular $X$. 
Therefore, {\tt HiggsBounds} must first determine $X_0$,
which is defined as the 
analysis application
with the highest statistical sensitivity
for the model point under consideration. In order to do this, the program
uses the tables of expected experimental limits to obtain a quantity
$Q_\EXPEC$ corresponding to each $X$. 
The analysis application with the largest value of
$Q_\MOD/Q_\EXPEC$ is chosen as $X_0$.
The program then 
determines
the value for $Q_\OBS(X_0)$,
using the appropriate 
observed limit. 
If 
${Q_\MOD(X_0)}{Q_\OBS(X_0)} > 1$,
{\tt HiggsBounds} concludes that this particular parameter point is excluded
at 95 \% C.L.

In \HB\ \vers, 82 Higgs search analyses have been implemented
(29 from LEP and 53 from the Tevatron). 
With respect to \HB\ {\tt 1}.{\tt 2}.{\tt 0} \cite{HB120-proceedings},
many new types of analyses have been added and 
several Tevatron analyses 
have been replaced by updated ones.
The following search topologies are implemented in \HB\ \vers.
(A full list of analyses with references is given in the manual
\cite{webpage}.)\smallskip

\noindent {$\bullet$ LEP neutral Higgs analyses considering the final states:}\\
$h_k Z, h_k \to b b$ %
/ $\tau \tau$ %
/ anything %
/ invisible %
/ $\gamma \gamma$ %
/ hadrons; %
\hspace{1mm}%
$b\bar b h_k \to b\bar b b\bar b$ / $b\bar b \tau\tau$ ($h_k$ CP even or odd)
	; %
\hspace{1mm}%
$\tau \tau h_k \to \tau\tau\tau\tau$ ($h_k$ CP even or odd)
	; %
\hspace{1mm}
$h_k Z, h_k \to h_i h_i, h_i \to b b$ / $\tau \tau$ ; %
\hspace{1mm}%
$h_k h_i,\;  h_k, h_i \to b b$ /$\tau \tau$ ; %
\hspace{1mm}%
\hspace{1mm}%
 $h_k h_i,\; h_k \to h_i h_i,\; h_i \to b b$ / $\tau \tau$ ; %
\hspace{1mm}%
\hspace{1mm}%
 $h_k Z, h_k \to h_i h_i, h_i \to b b, \tau \tau$
	; %
\hspace{1mm}%
 $h_k h_i,\; h_k \to b b, h_i \to \tau \tau$. %

\noindent {$\bullet$ Tevatron neutral Higgs  single topology 
	considering the final states:}\\
$Z h_k \to l l b \bar b$ 
	 ;\hspace{1mm}
 $W h_k \to l \nu b \bar b$
	;\hspace{1mm}
 $b h_k \to 3 b \text{ jets}$
	 ;\hspace{1mm}
 $\text{single } h_k \to W W$
	 ;\hspace{1mm}
 $\text{single } h_k \to \tau \tau$ %
	 ;\hspace{1mm}
 $W h_k \to 3 W$ 
	 ;\hspace{1mm} 
 $b h_k \to b \tau \tau$
	 ;\hspace{1mm}
 $t\bar th_k \to t\bar tb\bar b$
	 ;\hspace{1mm}
 $\text{single }h_k \to Z \gamma$.

\noindent {$\bullet$ Tevatron SM Higgs combined topologies 
	considering the final states (schematic):}\\
 $V h_k \to b \bar b + E_T^{\text{miss}} (V=W,Z)$ 
	; %
\hspace{1mm}
 $h_k + X \to W W + X$ 
	; %
\hspace{1mm}
 $h_k \to W W \to l l$
	; %
\hspace{1mm}
 $h_k + X \to \tau \tau$ 
	; %
\hspace{1mm}
 $h_k + X \to b b + X$  
	; %
\hspace{1mm}
 $Vh_k \to VVV \to l^\pm l^\pm + X$ ($l = e,\mu$, $V=W,Z$)  
	; %
\hspace{1mm}
 $h_k + X \to \gamma \gamma + X$;
\hspace{1mm}
 $h_k + X$.

\noindent {$\bullet$ LEP/Tevatron charged Higgs analyses:} \\
$e^+ e^- \to H^+ H^- \to$ 4 jets 
	; %
\hspace{1mm}
$e^+ e^- \to H^+ H^- \to \tau\nu\tau\nu$ 
	; %
\hspace{1mm}
$p\bar p\to tt, t \to H^+ b,\; H^+ \to cs$ (\& c.c.)
	; %
\hspace{1mm}
$p\bar p\to tt, t \to H^+ b,\; H^+ \to \tau \nu$ (\& c.c.).

\section{%
	 Applications}

\begin{figure}[t]
\includegraphics[width=.33\textwidth]{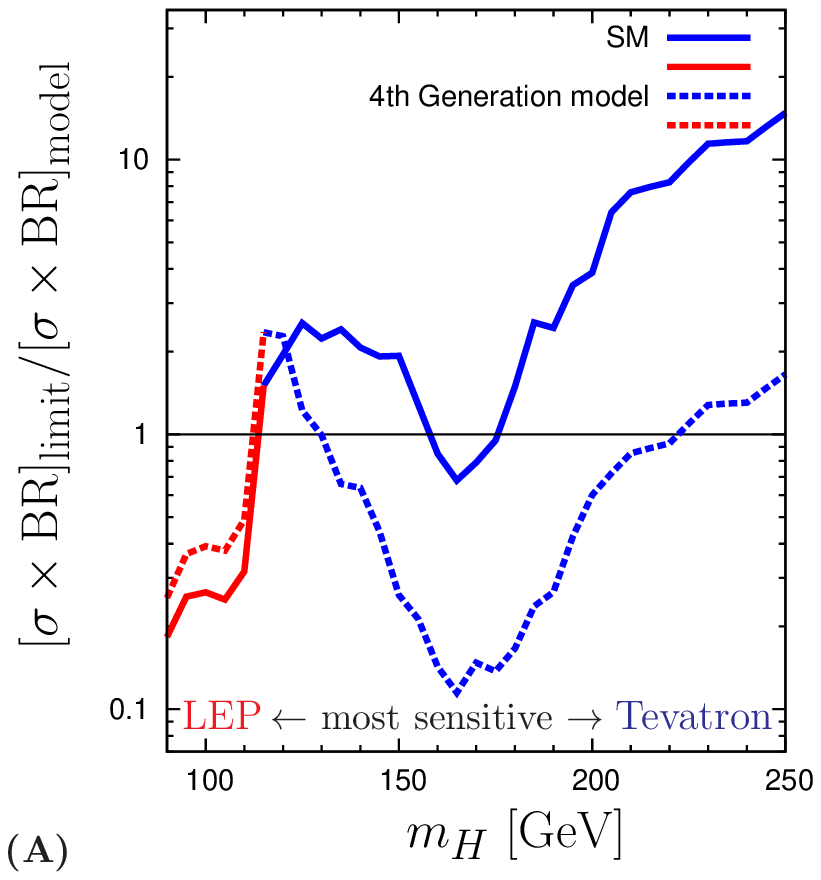}
\includegraphics[width=.66\textwidth]{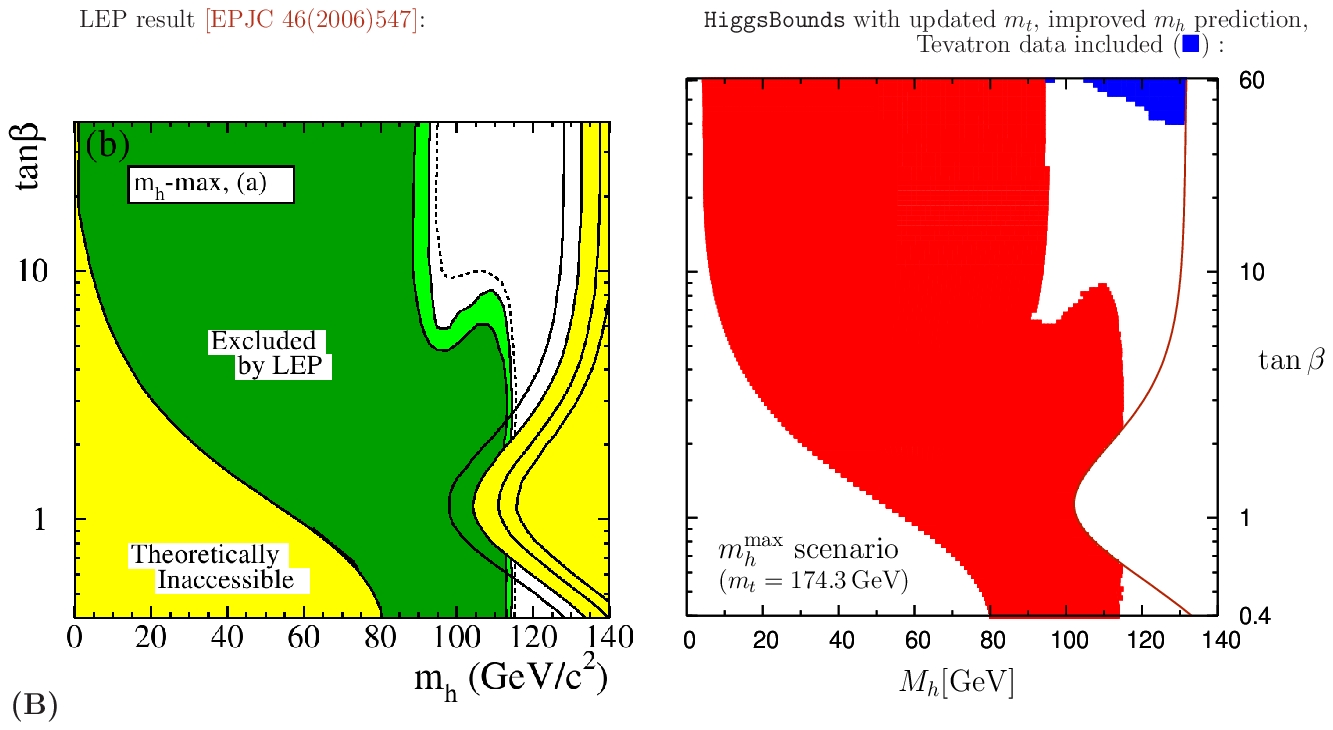}
\caption{\label{fig2} (A) 
Ratio $[\sigma\times \Br]_{\text{obs. limit }}/[\sigma\times \Br]_\MOD$ for the most sensitive
        analysis as a function of the Higgs mass $m_H$:
        4th Generation Model versus SM.
	(B) Excluded region in the $m_h-\tan\beta$ plane for 
	the MSSM $m_h^\MAX$ scenario. Results from
	LEP \cite{LEP-MSSM-Higgs-analysis} and obtained
	by using \HB\ \firstvers\ are compared.
}
\end{figure}

\noindent{\bf $\bullet$ 4$^{\text{th}}$ Generation Model (FGM)}:
As a toy example, we study a simplified FGM,
where we fix $\Gamma(H\to gg)_\MOD = 9\, \Gamma(H\to gg)_\SM$
for all Higgs masses%
\footnote{Assuming very heavy 4$^{\text{th}}$ generation quarks, 
an enhancement factor of 9 is only valid for $m_H \ll 2 m_t$,
i.e. towards large $m_H$
in Fig.~\ref{fig2}(A), 
this overestimates the signal cross section by $20\%$.}
.
Fig.~\ref{fig2}(A) shows the ratio of the signal cross section
limit vs. the model prediction for the SM (solid) and the 
FGM (dashed) obtained with \HB\ \vers.
The SM curve reproduces nicely the LEP and Tevatron exclusions
presented at the ICHEP'10.
In the FGM, 
$\sigma(p\bar p\to gg\to H\to W^+W^-)$ is strongly enhanced 
compared to the SM. Hence, the Tevatron Higgs mass exclusion 
in this model stretches from 130 GeV to 220 GeV.
Given our simplified approach, this is in very good agreement
with \cite{hep-ex/1005.3216}.

\begin{figure}[ht]
\includegraphics[width=1.0\textwidth]{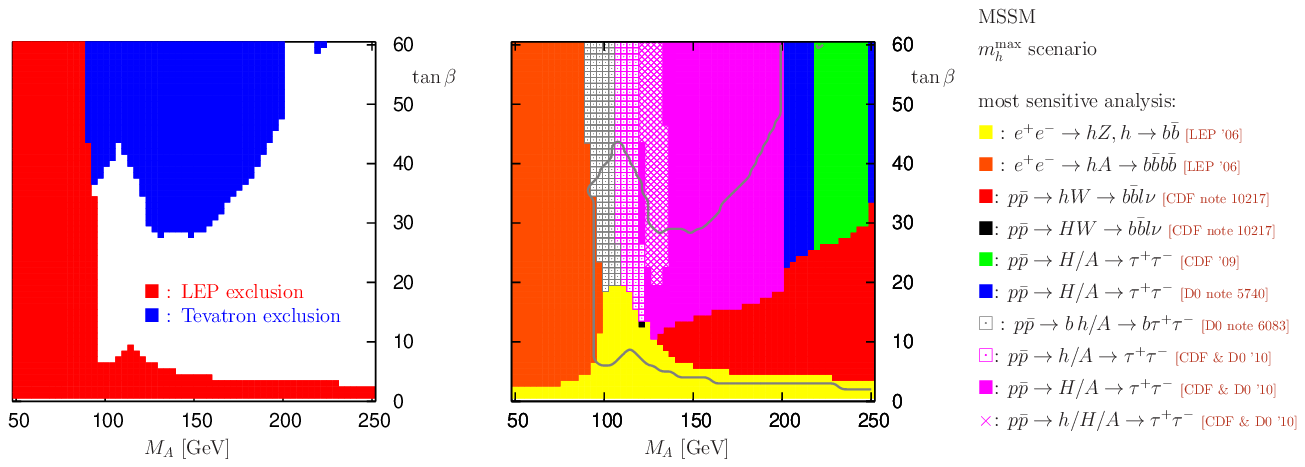}
\caption{\label{fig3} 
	Excluded region in the $m_A-\tan\beta$ plane for 
	the MSSM $m_h^\MAX$ scenario (left panel)
	and most sensitive analysis (right panel)
	using \HB\ \vers.
	}
\end{figure}

\noindent{\bf $\bullet$ MSSM Benchmark Scenarios:}
\HB\ is 
useful to 
update LEP exclusion plots for MSSM benchmark scenarios.
Fig.~\ref{fig2}(B) shows the good agreement between LEP results for the 
$m_h^\MAX$ scenario \cite{LEP-MSSM-Higgs-analysis} (left panel)
and results obtained with \HB\ {\tt 1}.{\tt 0}.{\tt 0} (right panel)
with updated MSSM predictions 
(using {\tt FeynHiggs} {\tt 2}.{\tt 6}.{\tt 4} \cite{FH}).
Verison 1.0.0 already contains all relevant LEP search
channels for this comparison.
Slight differences originate from the fact that
\HB\ relies on LEP data for single topological cross sections,
while the full LEP analysis combines the different search channels.
Fig.~\ref{fig3} shows for the MSSM $m_h^\MAX$ scenario the exclusion 
region (left panel) and the analyses with highest statistical sensitivity
in the $m_A-\tan\beta$ plane obtained with \HB\ \vers, 
i.e. including all Tevatron analyses presented at the ICHEP'10,
and {\tt FeynHiggs} {\tt 2}.{\tt 7}.{\tt 1} \cite{FH}
for the model predictions.
The gap in the Tevatron exclusion for 
$200\,\gev < m_A < 220\,\gev$
could possibly be filled if the 
CDF \& D\O\ combined $\tau\tau$
analysis \cite{hep-ex/1003.3363}, which has the
highest sensitivity for the Tevatron exclusion 
for 
$110\,\gev < m_A < 200\,\gev$, could be extended to
values $m_A > 200$ GeV.
\smallskip

\begin{figure}[ht]
\includegraphics[width=1.0\textwidth]{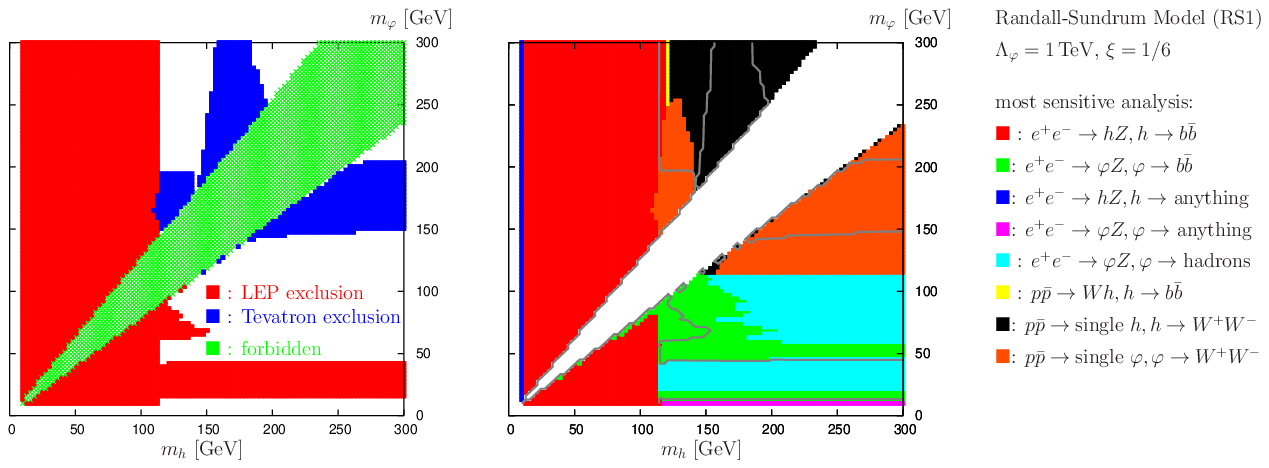}
\caption{\label{fig4} Excluded region in the $m_h-m_\varphi$ plane
	of the scalar sector of the original Randall-Sundrum model (RS1) 
	(left panel)
	and most sensitive search channel (right panel)
	using \HB\ \vers.
}
\end{figure}

\noindent{\bf $\bullet$ Scalar Sector of the Randall-Sundrum Model:}
The Randall-Sundrum (RS) Model considers spacetime a slice of 5d
anti-de-Sitter space with two boundaries, the IR brane
(our 4d spacetime) and the UV brane \cite{RS}, 
which naturally explains the hierarchy problem.
As a consequence of stabilising the compactification ``radius''
in the model, an additional scalar, 
the radion $\varphi_0$, appears in the spectrum \cite{GW}. 
Higgs--radion mixing may occur via the interaction
$
{\cal L}  = -\xi \sqrt{-g_{\text{ind}}}\: R(g_{\text{ind}})\: \Phi^\dagger \Phi
$,
where $g_{\text{ind}}$ is the induced 4d metric on the IR brane,
$R$ the Ricci scalar, and $\Phi$ the Higgs doublet.
Hence, in general,
the physical Higgs {$h_0$} and the radion {$\varphi_0$}
mix to form two mass eigenstates $h,\varphi$.
Like for the Higgs, the radion couplings to massive fermions
and gauge bosons are $\propto$ mass, but e.g. the couplings
$\varphi_0\, b \bar b$ and $\varphi_0\, \gamma\gamma$ are suppressed
while $\varphi_0\, g g$ is enhanced 
with respect to the SM Higgs. %
Although presently many modifications to the original RS Model (RS1) 
are considered, 
the RS scalar sector is a rather robust prediction,
as it is closely linked to the solution of the hierarchy problem.
Fig.~\ref{fig4} shows for RS1 the excluded region (left panel)
and the search channel with highest sensitivity (right panel)
in the $m_h-m_\varphi$ plane, while the two other free parameters 
in this model are set to $\Lambda_\varphi = 1\,\tev$ and $\xi = 1/6$.
This scenario shows slight Higgs--radion mixing, i.e. $h$ behaves 
mainly like the SM Higgs and $\varphi$ mainly like the unmixed radion. 
Given the properties of $h_0$ and $\varphi_0$, 
this is reflected in several ways by the pattern of LEP/Tevatron exclusions
and highest-sensitivity channels as a function of $m_h$ and $m_\varphi$.

\section{Summary}

The Higgs search at Tevatron and LEP turn(ed) out many limits
	on cross sections of individual and combined signal topologies.
Published individually, it is no easy to make use of all of them.
\HB\ \vers\ is a model-independent tool which
	offers easy access to a wealth of published limits
	in three ways: command line, subroutines, web interface.
	It offers a flexible range of input formats
	for the necessary model predictions, including the number of neutral
	{\em and charged} Higgs bosons.
Applications to a $4^{\text{th}}$ generation model, the MSSM and the
Randall-Sundrum model demonstrate its usefulness.
The code 
is publicly available \cite{webpage}.


\end{document}